\documentclass[12pt,letterpaper,journal,comsoc]{IEEEtran}
\usepackage[T1]{fontenc}      % modern font encoding
\usepackage{newtxtext,newtxmath} % Times-compatible text + math

\usepackage[margin=0.75in]{geometry} % generous side margins (≈½ in each side)
\usepackage{amsmath,amsfonts}
\usepackage{algorithmic}
\usepackage{algorithm}
\usepackage{array}
\usepackage[caption=false,font=normalsize,labelfont=sf,textfont=sf]{subfig}
\usepackage{textcomp}
\usepackage{stfloats}
\usepackage{url}
\usepackage{verbatim}
\usepackage{etoolbox} % for patching environments
\usepackage{cite}
\usepackage[utf8]{inputenc}
\usepackage{graphicx}
\usepackage{hyperref}
\usepackage{enumitem}
\usepackage[symbol]{footmisc}
\usepackage{lineno}
\usepackage{booktabs}
\usepackage{array}
\usepackage {dsfont}
\usepackage{algorithmic}
\usepackage{algorithm}
\usepackage{float}
\usepackage{times}

\usepackage{amssymb}
\DeclareUnicodeCharacter{221E}{\infty}  % Maps ∞ to \infty
\usepackage{amsfonts}
\usepackage{makeidx}
\usepackage{color}
\usepackage{comment}
\usepackage{listings}
\usepackage{xcolor}
\usepackage{array,multirow,graphicx}
\usepackage{fancyvrb}
\usepackage{tikz}
\usetikzlibrary{shapes.geometric, arrows, positioning}

\DefineVerbatimEnvironment{Prompt}{Verbatim}{
    fontfamily=cmtt,
    fontseries=b,
    frame=single,
    label=Prompt,
    labelposition=topline,
    breaklines=true,
    fontsize=\small  % Adjust font size if necessary
}

\DefineVerbatimEnvironment{LLMResponse}{Verbatim}{
    fontfamily=cmtt,
    frame=single,
    label=LLM Response,
    labelposition=topline,
    breaklines=true,
    fontsize=\small  % Adjust font size if necessary
}

\definecolor{codegreen}{rgb}{0,0.6,0}
\definecolor{codegray}{rgb}{0.5,0.5,0.5}
\definecolor{codepurple}{rgb}{0.58,0,0.82}
\definecolor{backcolour}{rgb}{0.95,0.95,0.92}

\lstdefinestyle{mystyle}{
  backgroundcolor=\color{backcolour}, commentstyle=\color{codegreen},
  keywordstyle=\color{magenta},
  numberstyle=\tiny\color{codegray},
  stringstyle=\color{codepurple},
  basicstyle=\ttfamily\footnotesize,
  breakatwhitespace=false,         
  breaklines=true,                 
  captionpos=b,                    
  keepspaces=true,                 
  numbers=left,                    
  numbersep=5pt,                  
  showspaces=false,                
  showstringspaces=false,
  showtabs=false,                  
  tabsize=2
}

\lstdefinestyle{yamlstyle}{
    basicstyle=\ttfamily\small,
    backgroundcolor=\color{gray!10},
    showspaces=false,
    showstringspaces=false,
    showtabs=false,
    tabsize=2,
    captionpos=b,
    breaklines=true,
    breakatwhitespace=true,
    breakautoindent=true,
    linewidth=\textwidth
}

\begin{document}

\title{Concept-Level AI for Telecom: Moving Beyond Large Language Models}

\author{
\IEEEauthorblockN{
Viswanath Kumarskandpriya\IEEEauthorrefmark{1}\thanks{\IEEEauthorrefmark{1}viswa.kumar@esme.fr},
Abdulhalim Dandoush\IEEEauthorrefmark{2}\thanks{\IEEEauthorrefmark{2}abdulhalim.dandoush@udst.edu.qa (Corresponding author)},
Abbas Bradai\IEEEauthorrefmark{2}\thanks{abbas.bradai@udst.edu.qa},
Ali Belgacem\IEEEauthorrefmark{3}\thanks{ali.belgacem@univ-cotedazur.fr}
}

\IEEEauthorblockA{\IEEEauthorrefmark{1}Esme Research Lab, SA ESME, Ivry-Sur-Seine, France} \\
\IEEEauthorblockA{\IEEEauthorrefmark{2}University of Doha for Science and Technology (UDST), Doha, Qatar} \\
\IEEEauthorblockA{\IEEEauthorrefmark{3}Côte d'Azur University, LEAT, Sophia Antipolis, France}
}

\maketitle

\begin{abstract}

The telecommunications and networking domain stands at the precipice of a transformative era, driven by the necessity to manage increasingly complex, hierarchical, multi administrative domains (i.e., several operators on the same path) and multilingual systems. Recent research has demonstrated that Large Language Models (LLMs), with their exceptional general-purpose text analysis and code generation capabilities, can be effectively applied to certain telecom problems (e.g., auto-configuration of data plan to meet certain application requirements). However, due to their inherent token-by-token processing and limited capacity for maintaining extended context, LLMs struggle to fulfill telecom-specific requirements such as cross-layer dependency cascades (i.e., over OSI), temporal-spatial fault correlation, and real-time distributed coordination. In contrast, Large Concept Models (LCMs), which reason at the abstraction level of semantic concepts rather than individual lexical tokens, offer a fundamentally superior approach for addressing these telecom challenges. By employing hyperbolic latent spaces for hierarchical representation and encapsulating complex multi-layered network interactions within concise concept embeddings, LCMs overcome critical shortcomings of LLMs in terms of memory efficiency, cross-layer correlation, and native multimodal integration. This paper argues that adopting LCMs is not simply an incremental step, but a necessary evolutionary leap toward achieving robust and effective AI-driven telecom management.

\end{abstract}

\section{Introduction}

Modern telecom networks feature layered architectures (for example, OSI model), distributed control planes, and multilingual environments, generating vast structured and unstructured data, from 3GPP documents to real-time logs and alarms. Traditional AI, including LLMs, struggles with this data due to three inherent limitations \cite{han2024lminfinite}:
\begin{itemize}
\item \emph{Token-centric processing}: LLMs fragment technical documents and logs into tokens, losing semantic relationships across protocol layers \cite{wang2024tokenization}, e.g., linking radio-link alarms to core-network states.
\item \emph{Memory constraints}: Correlating events across temporal or spatial dimensions (e.g., fault propagation, RAN-core interactions) exceeds LLMs' fixed attention window capabilities \cite{zhang2024inftybench}.
\item \emph{Multimodal rigidity}: Telecom data includes text (RFCs), speech (support calls), and structured signals (SNMP traps). Multi-modal LLMs often normalize non-text data into text, diluting meaning and adding latency \cite{yen2024multimodal}.
\end{itemize}

\begin{figure*}[ht]
     	\centering
     	\includegraphics [scale=0.425]{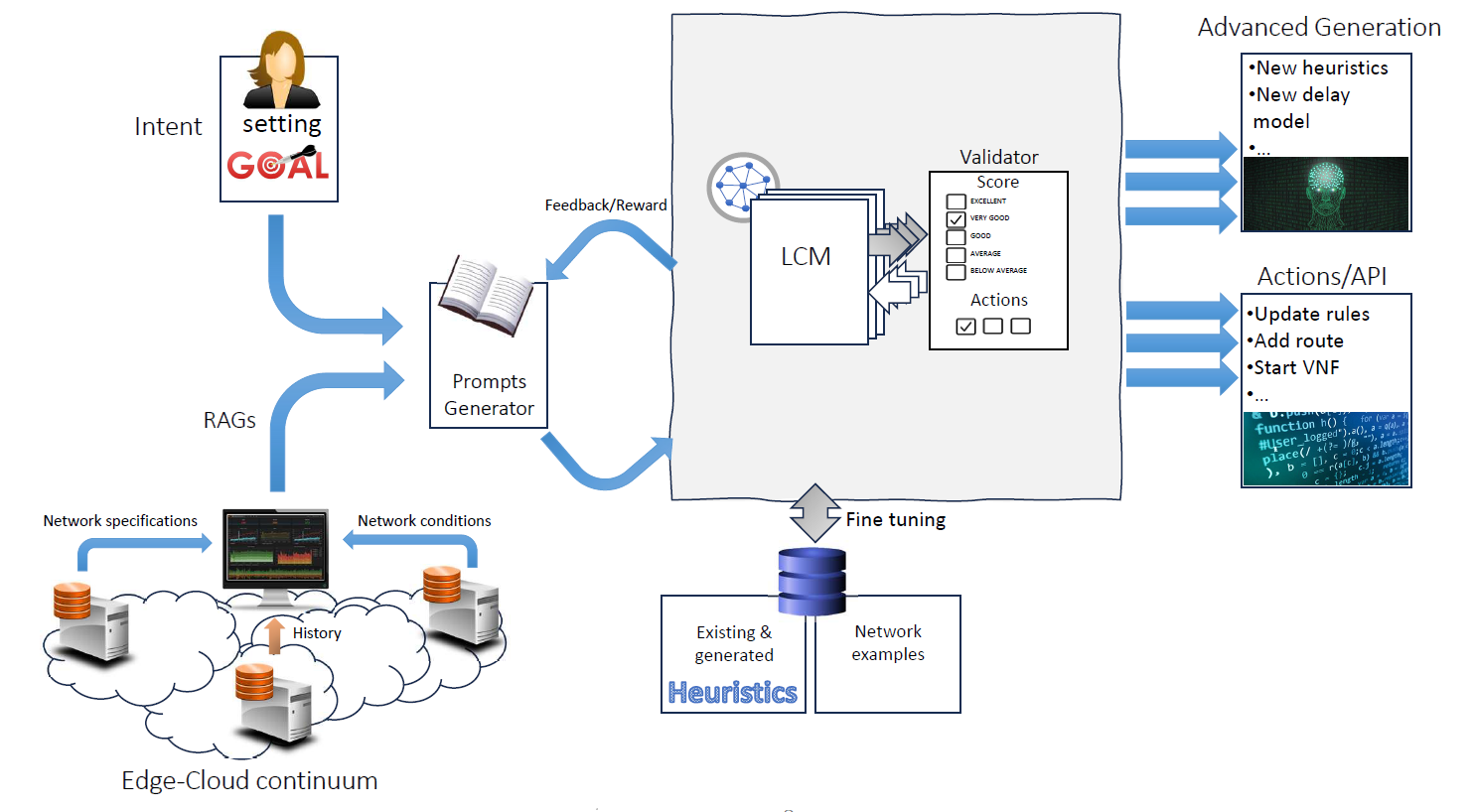} 
     	\caption  {Example on Closed-Loop Network Slice deployment Using Large Concept Models}
     	\label{bigpicture}
\end{figure*}

LCMs overcome these gaps via concept-level abstraction, enabling hierarchical reasoning and efficient knowledge compression. For example, a 5G network slice configuration—combining QoS, VLAN mappings, and user policies—can be represented as a single concept embedding instead of many disjoint tokens. This aligns naturally with telecom operations, where higher-layer services (e.g., VoLTE) abstract lower-layer complexities. Figure \ref{bigpicture} illustrates concept-level abstraction in an end-to-end control loop. An intent like ``instantiate an eMBB slice with 10 ms latency and $99.99\%$ availability'' augmented by live telemetry (e.g., link states, available compute power) via Retrieval-Augmented Generation (RAG) creates compact prompts querying a telecom-fine-tuned LCM. In fact, RAG enhances GenAI by combining it with external retrieval systems, ensuring up-to-date domain specific knowledge. The LCM reasons hierarchically over embedded concepts (e.g., QoS, VLAN tags, UE groups), producing optimized heuristics or executable actions (e.g., rule updates, VNF instantiation). A built-in validator ensures only high-quality outputs deploy, continuously refining the model using performance metrics.

\subsection{Overview of LLMs}

Large Language Models (LLMs) use in fact transformer-based neural networks with billions of parameters and large context windows \cite{vaswani2017attention}. These models utilize self-attention mechanisms, enabling each word to dynamically evaluate its relationships with others, uncovering dependencies between tokens. Also, Multi-head attention \cite{vaswani2017attention} further enhances this by simultaneously examining various word relationships, such as grammatical roles, semantic meanings, and syntax, helping decode intricate language patterns.

In addition, during pre-training, LLMs process diverse texts (books, articles, code) to learn language patterns by predicting sequential words, adjusting parameters to grasp grammar, logic, and domain-specific content like wireless communications. Their effectiveness arises from massive datasets and complex architectures, supporting extensive knowledge integration and sophisticated reasoning.

\subsubsection{Tokenization and Embeddings: Foundations of LLMs}

Tokenization segments raw text into tokens (words, subwords, characters) using algorithms like Byte Pair Encoding or WordPiece \cite{sennrich2016subword}, converting text into numerical sequences essential for computational processing. Embeddings transform tokens into dense, high-dimensional vectors encoding semantic and syntactic properties. These learned vectors position related meanings closer, allowing the model to leverage semantic similarities effectively. Thus, together, tokenization and embeddings underpin LLMs’ ability to comprehend, reason, and generate language fluently. In fact, without these foundational processes, sophisticated natural language handling by modern LLMs would be impossible.

\subsection{Overview of Large Concept Model}

The Meta's LCM paradigm \cite{lcmteam2024largeconceptmodelslanguage} elevates the atomic unit from \emph{token} to \emph{concept}. LCMs utilize concept encoders to map entire sentences or higher-level semantic units into a shared language-agnostic embedding space called SONAR (Sentence-Level Multimodal and Language-Agnostic Representations), which supports over 200 languages and both text and speech modalities. This architecture enables the model to reason and generate content in terms of concepts, aligning more closely with human abstraction and cognition.
The LCM architecture consists of a concept encoder that produces sentence-level embeddings, a transformer-based decoder that auto-regressively predicts sequences in this embedding space, and a concept decoder for reconstructing text or speech from embeddings. The model is trained using large-scale multilingual and multimodal data, and explores various generation strategies, including regression and diffusion-based methods.
Experimental results from the paper, show that LCMs outperform traditional LLMs of similar size in tasks requiring long-context reasoning, summarization, and cross-lingual generalization. By modeling language at the concept level, LCMs achieve greater coherence, interpretability, and efficiency, marking a significant advancement in the development of more robust and generalizable AI systems.

\subsection{LCMs vs LLMs}

%LLMs and LCMs differ fundamentally in their units of processing and the geometry of their representation spaces as shown in Figure \ref{lcmllm}. LCMs can be seen as one level of abstraction higher than LLMs. LLMs operate at the token level, breaking down text into words or sub-word units and generating language by predicting the next token in a sequence. This approach, rooted in Euclidean embedding spaces, excels at capturing local dependencies and producing fluent, grammatically correct text but often struggles with maintaining semantic coherence over long contexts and with abstract reasoning. The token-based nature of LLMs also leads to limitations in handling multilingual and multimodal content, as their representations are closely tied to the structure and vocabulary of specific languages. In contrast, LCMs process language at the level of concepts, which typically correspond to entire sentences or higher-level semantic units. These concepts are encoded into a high-dimensional, language and modality-agnostic embedding space, often hyperbolic in nature \cite{nickel2017poincare}, using specialized concept encoders such as SONAR. This design enables LCMs to reason over abstract ideas and maintain hierarchical relationships, leading to better performance in tasks requiring long-range context, cross-lingual generalization, and multimodal integration. By focusing on sentence-level or conceptual units, LCMs achieve greater semantic coherence, interpretability, and efficiency, marking a significant departure from the token-by-token paradigm of LLMs.

\begin{figure*}[ht]
     	\centering
     	\includegraphics [scale=0.145] {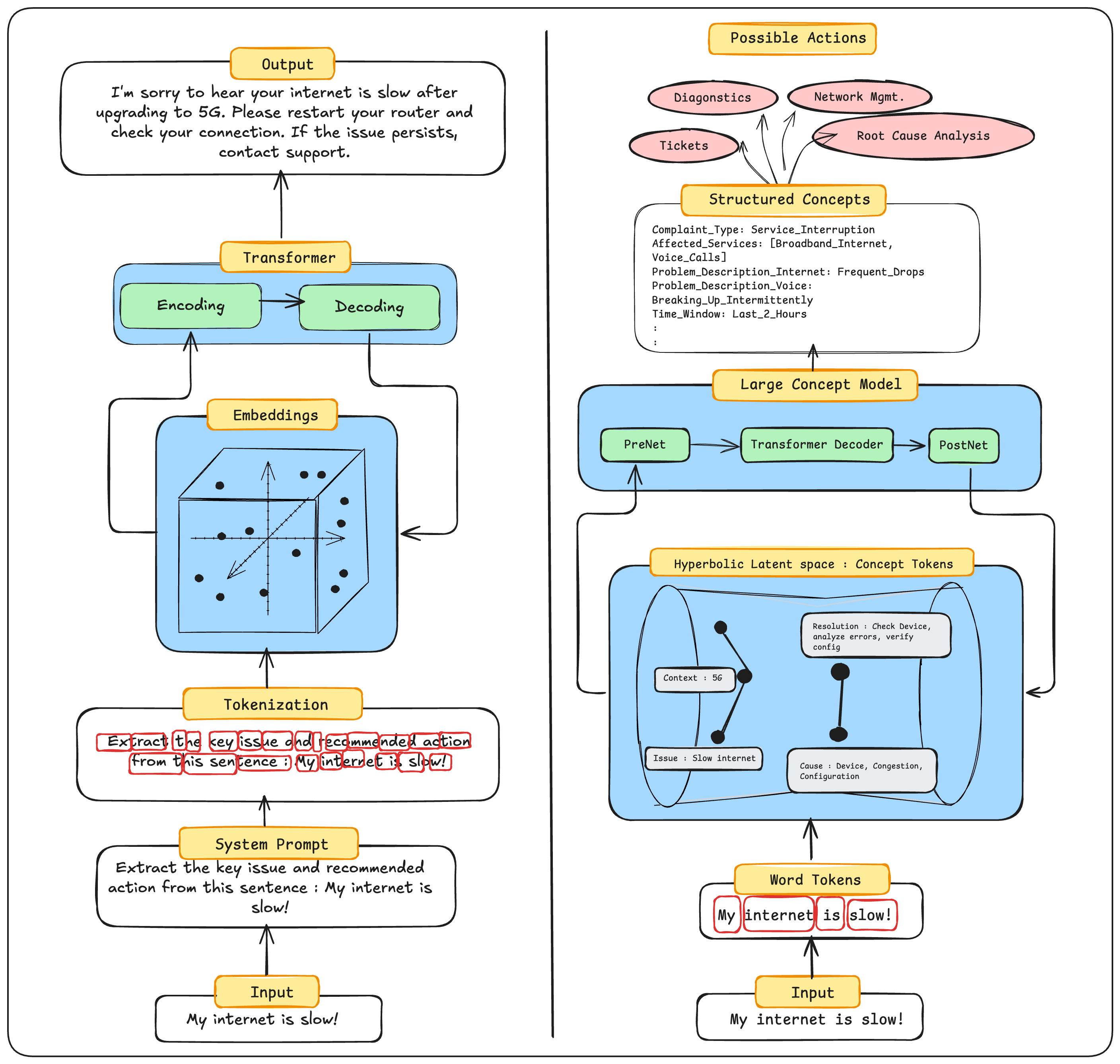}
     	%\caption  {Left: An input prompt with the help of system prompt is being consumed as tokens by LLM. These tokens are then represented as embeddings in euclidean space. LLM then use this representation to find most probable next token.         Right: An input prompt is represented as word tokens which is then presented as abstract concept tokens in hyperbolic latent space. LCM reasons on abstract concept level which is then converted back to needed format for post processing.}
        \caption{Comparison of LLM and LCM processing pipelines.}
     	\label{lcmllm}
\end{figure*}

LLMs and LCMs differ fundamentally in both (i) their processing granularity and (ii) the geometry of their internal representation spaces, as illustrated in Figure~\ref{lcmllm}. As depicted on the left side of the figure, LLMs operate at the token level, decomposing input prompts (including instructions) into discrete lexical units called tokens. Each token is mapped into an Euclidean embedding space, and the LLM predicts the next token based on statistical relationships learned from extensive training data. While this token centric approach effectively capture local syntactic dependencies and generates not only fluent but grammatically correct text, it faces inherent limitations when maintaining semantic coherence across extended contexts, abstract hierarchical reasoning, or integrating multimodal inputs. These constraints are particularly evident in telecom scenarios, where data is multilingual, hierarchical, and multimodal in addition to the fac that we have different administrative domains using each a different modeling language, north and south bound interfaces, different templates of the configuration files to mention a few.

In contrast, as shown on the right side of Figure~\ref{lcmllm}, LCMs abstract the input tokens into higher-level semantic concepts, such as complete sentences or complex network configurations, rather than individual lexical tokens. These concepts are encoded into a high dimensional, language agnostic latent space, often hyperbolic in nature~\cite{nickel2017poincare}. Such hyperbolic embeddings naturally preserve hierarchical relationships and allow efficient reasoning over entire semantic units. Specialized encoders like Meta’s SONAR further facilitate this process by enabling the integration of multimodal data (text, speech, telemetry) into a unified conceptual representation. The conceptual reasoning capability of LCMs allows them to better manage long-range dependencies, multilingual content, and complex multimodal data, thereby offering significant advantages in semantic coherence, interpretability, and efficiency over token-by-token processing used by traditional LLMs.

\section{Related Works}

We compare recent works applying generative AI to telecom using LLMs. TSpec-LLM \cite{nikbakht2024tspecllmopensourcedatasetllm} introduced a dataset covering 3GPP documents from Release 8 to 19, significantly boosting GPT-3.5 and GPT-4 accuracy from below $51\%$ to above $71\%$ through Retrieval-Augmented Generation (RAG). \cite{lin2024primergenerativeaitelecom} proposed a practical framework highlighting RAG’s importance for connecting LLMs to telecom-specific knowledge, notably demonstrating an O-RAN chatbot gaining industry recognition and providing open-source implementations for real-world utility. Telco-RAG \cite{bornea2024telcoragnavigatingchallengesretrievalaugmented} specialized in addressing challenges of applying RAG to technical telecom documentation, particularly complex 3GPP standards, offering guidelines relevant to broader technical domains. TeleQnA \cite{maatouk2023teleqnabenchmarkdatasetassess} introduced the first benchmark with 10,000 telecom $Q\&A$ pairs, indicating advanced LLMs struggle with complex standards and showing context-enriched approaches substantially enhance performance. The dataset was created via automated generation with human verification. \cite{bariah2023largegenerativeaimodels} fine-tuned BERT, RoBERTa, and GPT-2 on telecom language from 3GPP documents, achieving $84.6\%$ accuracy in identifying working groups, proving domain-specific adaptation is effective even for smaller models. In \cite{dandoush2024largelanguagemodelsmeet} LLMs multi-agent systems for network slicing management is introduced. Scalability and interoperability were identified to be the key challenges for effective AI-driven orchestration in future networks. In \cite{bornea2024telcoragnavigatingchallengesretrievalaugmented} the authors envisioned GenAI transforming wireless networks into autonomous systems, using multi-modal models trained on diverse telecom data.

Previous works generally focus on enhancing LLM performance via curated benchmarks, dataset fine-tuning, or context augmentation (e.g., RAG), highlighting inherent LLM limitations like memory constraints and tokenization issues.

\section{Why LCMs Excel in Telecom and Networking Applications}

LLMs tokenize input streams like logs, alarms, and configurations into discrete subword units, causing a loss of semantic structure critical for telecom network analysis. For instance, identifying a high-level service disruption caused by a low-level anomaly often requires correlating events separated by thousands or millions of tokens. Constrained by fixed attention windows and token limits, LLMs struggle to retain context, leading to incomplete or incorrect fault analysis. Additionally, their text-centric nature limits integrating structured data, diagrams, and multimodal signals common in telecom, such as SNMP traps, configuration tables, and network topologies. LCMs overcome these limitations by operating on concept-level units, complete sentences or semantic units, encoded in a language-agnostic, often hyperbolic embedding space. This approach compresses extensive operational histories and cross-layer dependencies into meaningful concept tokens, maintaining the hierarchical relationships intrinsic to telecom systems. Practically, LCMs enable effective alarm correlation, root cause analysis, and log parsing by representing complex event chains as interconnected concepts rather than fragmented tokens. For example, an LCM can transform multi-layered, multimodal alarm and log sequences into coherent conceptual graphs, facilitating rapid root cause identification and actionable insights. This enhances interpretability, generalization, and multilingual/multimodal data handling, making LCMs particularly suited for telecom applications like alarm flood reduction, automated ticketing, and cross-domain fault correlation.

\section{ Case study: Intent-based Networking (IBN)}
The telecoms industry is undergoing a radical shift from traditional rule-based models to intelligent systems that understand users' intentions and automatically translate them into network actions. Intent is defined as the expression of a user's desired goal, such as improving service quality or optimising energy consumption, without the need for in-depth technical knowledge.
\par
To achieve this vision, AI techniques are used to analyse user intentions through natural language and integrate multi-source data (performance metrics, traffic patterns, spectrum resources). While Large Language Models (LLMs) offer advanced language understanding capabilities, they face challenges in this area, such as:
\begin{itemize}
    \item Lack of specialisation in protocols (5G NR, IMS, O-RAN)
    \item Difficulty in verifying outputs and ensuring regulatory compliance. LLMs may exaggerate their results or propose actions that violate key communications standards, such as 3GPP and ETSI, due to their multipurpose nature.
    \item High computing requirements that hinder real-time response
    \item These models require large, often irrelevant, datasets for fine-tuning, reducing their effectiveness in specialized communications environments.
    \item  LLMs lack strong causal and temporal inference capabilities, making it difficult to analyze patterns in network logs and KPIs.
    \item These models have difficulty integrating and interpreting various real-time data sources, such as telemetry, performance metrics, and configuration files.
\end{itemize}

\par
Language Concept Models (LCMs) are a specialized and efficient alternative, based on knowledge structures and conceptual schemas specifically designed to understand the communication environment. Thanks to their symbolic structure, LCMs can accurately interpret intentions and provide interpretable outputs, ensuring a balance between business goals and network behaviour. Moreover, LCMs are computationally lightweight and can adapt almost instantaneously to changes, making them more suitable for implementing real-time network policies. Unlike LLMs, LCMs consider regulatory rules and standards from design, ensuring that their outputs are compliant with standards such as 3GPP and ETSI (Table\ref{tab:llm-lcm-ibn}).
\par
\begin{figure}[h]
     	\centering
     	\includegraphics [scale=0.35]{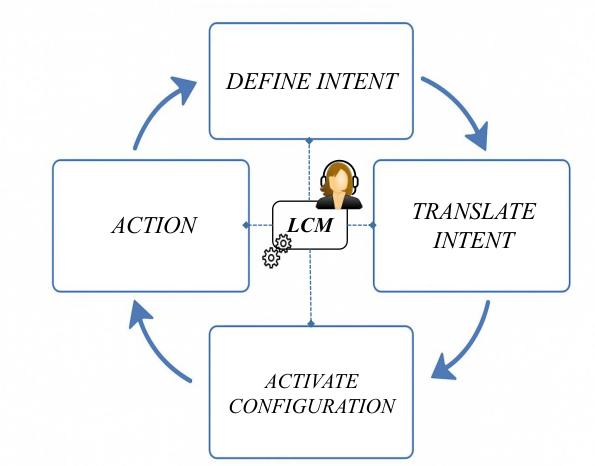} 
     	\caption  { LCM-Driven Stages in IBN}
     	\label{IBNLifCy}
\end{figure}

Integrating LCM modules into every stage of the network lifecycle transforms network management from a complex, manual process to a flexible experience that relies on human linguistic interaction (Figure \ref{IBNLifCy}). LCMs act as an intelligent intermediary between operators and network components, allowing for accurate interpretation and inference of intentions, and more efficient execution of tasks. %This approach not only reduces the need for specialized technical knowledge, but also enhances the network's flexibility and responsiveness to the rapid changes and demands of the modern telecommunications environment.

\begin{table*}[h]
\centering
\renewcommand{\arraystretch}{1.3}
\begin{tabular}{>{\bfseries}l p{4.5cm} p{5.5cm}}
\toprule
IBN Stage & LLM Focus & LCM Focus \\
\midrule
Define intent & Natural Language to Intent & Semantic Concept Extraction \\
Translate intent & Policy Rule Generation & Concept Mapping \& Validation \\
Activate Configuration & Script/Config Generation & Goal-Compliance \& Semantic Alignment \\
Action & Interpret Output \& Feedback & Intent Assurance \& Concept Reasoning \\
\bottomrule
\end{tabular}
\caption{Roles of LLM and LCM in different IBN stages}
\label{tab:llm-lcm-ibn}
\end{table*}

\section{Challenges and Future directions}

The development of LCMs remains in its nascent stages, with the research community only beginning to explore their full potential and practical applications. Current implementations of concept encoders, such as Meta’s SONAR, represent pioneering efforts but are limited in availability and scope. This scarcity of mature, widely accessible concept encoding frameworks poses a significant barrier to advancing LCM research, particularly in specialized application domains like telecommunications and networking. The lack of standardized datasets annotated at the concept level further compounds this challenge, making it difficult to rigorously evaluate and benchmark LCM performance against traditional models or to tailor them effectively to domain-specific complexities.
Moreover, many foundational aspects of LCM architectures remain underexplored. For instance, the optimal geometric configurations for embedding hierarchical telecom data, the integration of multimodal signals beyond text and speech, and the development of scalable training algorithms that preserve concept-level semantics across evolving network environments are open research questions. Addressing these issues will require innovations in model design, such as hybrid embedding spaces, sparse attention mechanisms adapted to concept hierarchies, and domain-specific pretraining strategies that incorporate telecom standards and operational data.
Future research directions should focus on expanding the concept encoders to cover a broader range of telecom modalities and languages, fostering the creation of comprehensive, annotated datasets that capture the hierarchical and distributed nature of telecom systems. To that end, collaborative efforts between academia, industry consortia like GSMA, and network operators could accelerate the development of benchmarks and shared resources \cite{gsma2025opentelco}. Furthermore, exploring the interaction between LCMs and emerging hardware accelerators could unlock new efficiencies. %Ultimately, advancing LCMs will require a multidisciplinary approach that bridges machine learning theory, telecom domain expertise, and systems engineering to realize models capable of meeting the complex demands of next-generation networks.

\section{Conclusion}
We highlighted LCMs' unique suitability for telecommunications and networking, emphasizing advantages from their concept embedding spaces. Unlike LLMs, limited by tokenization, memory constraints, and text-centric architectures, LCMs encode whole sentences or semantic units as concepts in hyperbolic and language-agnostic embedding spaces. This shift naturally captures hierarchical, distributed, and multimodal telecom dependencies. The case study demonstrated LLM limitations which significantly challenge telecom applications. Conversely, LCMs' concept-level reasoning provides superior, coherent, and actionable insights. Last, we acknowledge LCM research remains nascent. Though implementations like SONAR pioneered concept-based modeling, robust concept encoders and domain-specific, concept-annotated datasets remain limited compared to mature LLM technologies.\\ %Advancing LCM research requires new architectures, expanded resources, and academia-industry collaboration. Nevertheless, the evidence presented strongly supports LCMs as transformative for telecom AI, potentially addressing longstanding challenges and enabling next-generation network capabilities.
\vspace{0.3em}
\textbf{Acknowledgment:} We used AI tools, e.g., ChatGPT-4, to assist with editing and grammar refinement in several sentences. The intellectual content, however, reflects the authors’ original contributions and expertise. Also, all authors made a significant intellectual contribution in this study, participated in the drafting, reviewing and approving the final manuscript.

\bibliographystyle{IEEEtran}
\bibliography{ref.bib}
\section*{Biographies}

\textbf{Abdulhalim Dandoush} is an asso. prof. and head of IT Dept at UDST, Qatar. He obtained a PhD from INRIA Nice-Sophia Antipolis and the university of Cote-d'Azur in France in 2010. He is leading the IT research team at ESME, France. He is actively working on Intelligent and self-driven Network Management.

\vspace{0.3em}
\textbf{Viswanath Kumarskandpriya} is an ICT Engineer and PhD candidate with over 17 years of experience in the telecom industry. He holds an engineering degree in electronics and communications and a masters in software engineering. His work focuses on routing, security, ML-based network optimization and Intelligent network management. Also, he is an active contributor to open source communities like ONOS.

\vspace{0.3em}
\textbf{Ali Belgacem} holds a PhD in CS and is working at LEAT laboratory in France. He served as an Asso. Prof. at the University of Boumerdes. He was also a postdoc researcher at the XLIM laboratory.  His research focuses on dynamic resource allocation in cloud/Edge computing, with additional interests in IoT, Artificial Intelligence (AI), LLMs and LCMs.

\vspace{0.3em}
\textbf{Abbas Bradai} is a Full Professor at UDST, Qatar, and former Full Professor at University of Cote d'Azur, France. With over 2,200 citations in IoT, SDN/NFV, and wireless networking, his research spans industrial applications and academic collaborations.

\end{document}